\documentclass[11pt]{article}

\usepackage{graphicx,amsmath,amsfonts,amssymb, dcolumn,amsthm}
\usepackage{amsmath,amssymb,amsfonts,mathrsfs}	
\usepackage{slashed}
\usepackage{bbm}                
\usepackage{xspace}
\usepackage{pgffor}	
\usepackage{tikz}
\usepackage{mathbbol}
\usepackage{latexsym}
\usepackage{bbm} 
\usepackage{slashed}            
\usepackage{bbm}                
\usepackage{xspace}				

\numberwithin{equation}{section} 
\usepackage{tikz}
\usetikzlibrary{arrows,shapes}
\usetikzlibrary{trees}
\usetikzlibrary{matrix,arrows} 				
\usetikzlibrary{positioning}				
\usetikzlibrary{calc,through}				
\usetikzlibrary{decorations.pathreplacing}  
\usepackage{pgffor}							

\usetikzlibrary{decorations.pathmorphing}	
\usetikzlibrary{decorations.markings}
\tikzset{
    vector/.style={decorate, decoration={snake}, draw},
	provector/.style={decorate, decoration={snake,amplitude=2.5pt}, draw},
	antivector/.style={decorate, decoration={snake,amplitude=-2.5pt}, draw},
    fermion/.style={draw=black, postaction={decorate},
        decoration={markings,mark=at position .55 with {\arrow[draw=black]{latex}}}},
    fermionbar/.style={draw=black, postaction={decorate},
        decoration={markings,mark=at position .55 with {\arrow[draw=black]{latex reversed}}}},
    fermionnoarrow/.style={draw=black},
    gluon/.style={decorate, draw=black,
        decoration={coil,amplitude=4pt, segment length=5pt}},
    scalar/.style={dashed,draw=black, postaction={decorate},
        decoration={markings,mark=at position .55 with {\arrow[draw=black]{latex}}}},
    scalarbar/.style={dashed,draw=black, postaction={decorate},
        decoration={markings,mark=at position .55 with {\arrow[draw=black]{latex}}}},
    scalarnoarrow/.style={dashed,draw=black},
    electron/.style={draw=black, postaction={decorate},
        decoration={markings,mark=at position .55 with {\arrow[draw=black]{latex}}}},
	bigvector/.style={decorate, decoration={snake,amplitude=4pt}, draw},
}

\tikzstyle{block} = [draw, rectangle, 
    minimum height=3em, minimum width=6em]

\begin{document}

\title{\textbf{Lorentz-violating Yang-Mills theory: discussing the Chern-Simons-like term generation}}
\author{ \textbf{Tiago R.~S.~Santos}\thanks{santostrs@gmail.com}\  \ and \textbf{Rodrigo F.~Sobreiro}\thanks{sobreiro@if.uff.br}\\\\
\textit{{\small UFF $-$ Universidade Federal Fluminense,}}\\
\textit{{\small Instituto de F\'{\i}sica, Campus da Praia Vermelha,}}\\
\textit{{\small Avenida General Milton Tavares de Souza s/n, 24210-346,}}\\
\textit{{\small Niter\'oi, RJ, Brasil.}}}
\date{}
\maketitle

\begin{abstract}
We analyze the Chern-Simons-like term generation in the CPT-odd Lorentz-violating Yang-Mills theory interacting with fermions. Moreover, we study the anomalies of this model as well as its quantum stability. The whole analysis is performed within the algebraic renormalization theory, which is independent of the renormalization scheme. In addition, all results are valid to all orders in perturbation theory. We find that the Chern-Simons-like term is not generated by radiative corrections, just like its Abelian version. Additionally, the model is also free of gauge anomalies and quantum stable.
\end{abstract}

\section{Introduction}\label{intro}

Many theoretical results have been obtained with respect to the renormalization aspects and radiative inductions of the minimal sector of the Standard Model Extension \cite{Colladay:1996iz,Colladay:1998fq,Jackiw:1999yp,Kostelecky:2003fs,Kostelecky:2005ic,Diaz:2011ia}. For instance, the renormalizability of Lorentz-violating QED was verified at one-loop order in \cite{Kostelecky:2001jc}. This result was generalized for curved manifold in \cite{deBerredoPeixoto:2006wz}. Moreover, gauge anomalies aspects and the all the orders renormalizability of this model also were verified \cite{Franco:2013rp,Vieira:2015fra,DelCima:2012gb,Santos:2016bqc,Santos:2015koa}. In the works \cite{DelCima:2012gb,Santos:2016bqc,Santos:2015koa}, the algebraic renormalization approach \cite{Piguet:1995er} was employed. The novelty introduced in Refs.~\cite{Santos:2016bqc,Santos:2015koa} is the introduction of the Symanzik method \cite{Symanzik:1969ek} and the Becchi-Rouet-Stora-Tyutin (BRST) quantization \cite{Becchi:1975nq,Tyutin:1975qk}. 

The issue of radiative induction of the Chern-Simons-like term in the Lorentz-violating QED was object of intense debate. For instance, by a non-perturbative analysis of Feynman integrals it is argued that the Chern-Simons-like term is generated by radiative corrections and is determined \cite{Jackiw:1999yp,Chung:1998jv,Chung:1999gg,Chung:1999pt,PerezVictoria:1999uh}. It is worth mentioning that in \cite{Jackiw:1999yp}, when the perturbative approach is employed, a Chern-Simons-like term is generated but it is ambiguous (If the Pauli-Villars regularization \cite{Pauli:1949zm} is employed, there is no generation of the Chern-Simons-like term). In Ref.~\cite{Chen:1999ws}, making use of the differential regularization \cite{Freedman:1991tk}, ambiguities also show up and the authors argue that such ambiguity should be fixed by some physical condition or some fundamental principle. In \cite{Chaichian:2000eh}, using the proper-time Schwinger method \cite{Schwinger:1951nm}, it is obtained the same result found through covariant derivative expansion \cite{Chan:1999nk}, but these results differ from results found from other regularization schemes. It is worth to mention that in \cite{Chan:1999nk} no regularization scheme is used; the point raised by the authors is that the use of an invariant regularization method for finite integrals (in order to keep the gauge symmetry of the theory) will avoid all anomalous terms. For instance, in Ref.~\cite{Jackiw:1999qq} it is claimed that whether the gauge symmetry is not used (transversality of the gauge field propagator in the Landau gauge), the ambiguity of the Chern-Simons-like term persists. In contrast, Ref.~\cite{PerezVictoria:2001ej} claims that the gauge symmetry does not fix the ambiguity. Nevertheless, in Refs.~\cite{Bonneau:2000ai,Bonneau:2006ma}, by Ward identities arguments, it is shown that the Chern-Simons-like term is not generated by radiative corrections, perturbatively or not. 

As we can note, the study of the possible generation of the Chern-Simons-like term under various different regularization schemes is source of confusion, leading to different answers. Perhaps, the final answer to the question whether the Chern-Simons term is generated by radiative corrections or not could arise from the algebraic renormalization point of view, a renormalization independent method. In fact, the authors in Refs.~\cite{Santos:2016bqc,Santos:2015koa,DelCima:2009ta}, using such method, have shown that a Chern-Simons-like term is not generated in the Lorentz-violating QED. 

In the case of non-Abelian Lorentz-violating models \cite{Colladay:2006rk}, the literature is quite poorest. To our knowledge, the problem of the radiative generation of the Chern-Simons-like term was only addressed in Ref.~\cite{Gomes:2007rv}, where the authors show that such a term is regularization scheme dependent for zero and finite temperatures. Thus, it is opportune to analyze this issue under another point of view. Besides, to know if the Chern-Simons-like term is generated or not is very important for new contributions on the mass terms in the Lorentz-violating Yang-Mills theory \cite{Santos:2014lfa}. In fact, as pointed in Ref.~\cite{Santos:2014lfa}, in contrast to the Abelian theory \cite{Santos:2015koa}, mass terms are generated from the bosonic CPT-odd sector. Such mass terms will modify drastically the gauge field propagator \cite{Santos:2016uds}. This could indicate that the presence of a Lorentz-violating sector in the Yang-Mills theory could affect the Gribov problem \cite{Gribov:1977wm} in this theory \cite{Granado:2017xjs}. Thus, the generation of a Chern-Simons-like term from fermionic CPT-odd sector would affect the physical spectrum of the model more than was pointed in Ref.~\cite{Santos:2016uds}.

For these reasons, in the present work we study the Chern-Simons-like term radiative generation under the algebraic renormalization theory. As said before, one of the main advantages of this technique is its regularization scheme independence \cite{Piguet:1995er, Piguet:1980nr,Lowenstein:1971vf,Clark:1976ym, Lam:1972mb,Duetsch:2000nh}. Following the prescriptions developed in \cite{Santos:2016bqc,Santos:2015koa,Santos:2014lfa}, we are able to control all symmetry violations through the Symanzik method \cite{Symanzik:1969ek} of external sources and BRST quantization\footnote{For further applications of the Symanzik method together with the BRST quantization we refer to  Refs.~\cite{Zwanziger:1992qr,Dudal:2005na,Baulieu:2008fy,Baulieu:2009xr,Dudal:2011gd,Pereira:2013aza}.} \cite{Piguet:1995er,Becchi:1975nq,Tyutin:1975qk}. Within this approach we are able not only to show that the Chern-Simons term is not radiatively generated but that the model is free of gauge anomalies as well as it is stable under quantum corrections. Moreover, the results here presented are valid to all orders in perturbation theory. 

We present this work as follows. In Sec.~\ref{YM} the Lorentz-violating Yang-Mills theory with interacting fermions is presented. The Sec.~\ref{QUATIZATION} treats the BRST quantization of the model in addition with the Symanzik procedure. In Sec.~\ref{GA}, we study the existence of gauge anomalies in the model by extending the Ward identities to the quantum level. The Sec.~\ref{QS} is devoted to the study of the quantum stability of the model (Is at this point that we show that the Chern-Simons-like term is not generated by radiative corrections). Then, in Sec.~\ref{FINAL} we present our final considerations.

\section{Lorentz-violating Yang-Mills theory}\label{YM}

As said before, we shall consider the Yang-Mills theory, for the $SU(N)$ symmetry group, including a term with Dirac fermions. The gauge fields are algebra-valued $A_{\mu}=A^a_{\mu}T^a$, where $T^a$ are the generators of the $SU(N)$ algebra, chosen to be anti-Hermitian and have vanishing trace and normalized as $\textrm{Tr}(T^aT^b)=\delta^{ab}$. The Lie algebra is given by $[T^a,T^b]=f^{abc}T^c$, where $f^{abc}$ are the skew-symmetric structure constants. The Latin indices run as $\left\{a,b,c,\dots\right\}\;\in\;\left\{1,2,\dots,N^2-1\right\}$. Furthermore, we add to this theory a Lorentz-violating sector following the mSME criteria. However, to avoid a cumbersome analysis, we consider here, just a sector of CPT-odd, both for bosonic and fermionic sectors.

With the prescription aforementioned, the model is described by the following action 
\begin{eqnarray}
\Sigma_{LV}&=&\Sigma_{YM}+\Sigma_D+\Sigma_{LVB}+\Sigma_{LVF}\;,
\label{YM0}
\end{eqnarray}
where
\begin{eqnarray}
\Sigma_{YM}&=&-\frac{1}{4}\int d^4x\;F^a_{\mu\nu}F^{a\mu\nu}\;
\label{YM1}
\end{eqnarray}
is the Yang-Mills action and 
\begin{eqnarray}
\Sigma_{D}&=&\int d^4x\;\overline{\psi}(i\gamma^{\mu}D_{\mu}-m)\psi\;
\label{YM2}
\end{eqnarray}
is the Dirac action\footnote{To avoid a cumbersome notation, we have omitted here the internal index of fundamental representation of $SU(N)$, i.e., $\psi\equiv\psi^i$ and $T^a\equiv(T^a)^{ij}$.}. The field strength is defined as $F^a_{\mu\nu}\equiv\partial_{\mu}A^a_{\nu}-\partial_{\nu}A^a_{\mu}+gf^{abc}A^b_{\mu}A^c_{\nu}$. The covariant derivative in the fundamental representation is defined as $D_{\mu}\equiv\partial_{\mu}+gA^a_{\mu}T^a$. $\psi$ is the Dirac field, and its Dirac adjoint is denoted by $\overline{\psi}=\psi^{\dag}\gamma^{0}$. The parameter $m$ stands for the electron mass and $g$ for the Yang-Mills coupling parameter. The $\gamma^\mu$ matrices are in Dirac representation. The bosonic Lorentz-violating sector of CPT-odd is described by the following action
\begin{eqnarray}
\Sigma_{LVB}&=&\int d^4x\;\epsilon_{\mu\nu\alpha\beta}v^{\mu}\left(A^{a\nu}\partial^{\alpha}A^{a\beta}+\frac{g}{3}f^{abc}A^{a\nu}A^{b\alpha}A^{c\beta}\right)\;,
\label{YM3}
\end{eqnarray}
and, for our proposes, we consider just one Lorentz-violating term of CPT-odd in the fermionic sector\footnote{This is the fermionic term that could give rise to the Chern-Simons-like term, i.e., at one-loop order we might have $v^\mu =\zeta \kappa^\mu$, where $\zeta$ is a parameter depending on the coupling parameter.}, namely,
\begin{eqnarray}
\Sigma_{LVF}&=&-\int d^4x\;\kappa^{\mu}\overline{\psi}\gamma_5\gamma_{\mu}\psi\;.
\label{YM4}
\end{eqnarray}
The violation of Lorentz symmetry in the bosonic and fermionic sectors is characterized by the constant vectors $v^{\mu}$ and $\kappa^{\mu}$, respectively. Both violating parameters carry mass dimension 1.

\section{The BRST quantization and the Symanzik approach} \label{QUATIZATION}

\subsection{BRST gauge fixing}

In order to quantize a gauge theory to obtain a consistent gauge field propagator \cite{Santos:2016uds} a gauge fixing is needed. For simplicity, we choose the Landau gauge condition, i.e., $\partial_\mu A^{a\mu}=0$. The BRST quantization method will be employed. Therefore, we introduce the Lautrup-Nakanishi field $b^a$ and the Faddeev-Popov ghost and antighost fields, namely, $c^a$ and $\overline{c}^a$, respectively. The BRST transformations of the fields are:
\begin{eqnarray}
sA^a_{\mu}&=&-D^{ab}_{\mu}c^b\;,\nonumber\\
sc^a&=&\frac{g}{2}f^{abc}c^bc^c\;,\nonumber\\
s\psi&=&gc^aT^a\psi\;,\nonumber\\
s\overline{\psi}&=&g\overline{\psi}c^aT^a\;,\nonumber\\
s\overline{c}^a&=&b^a\;,\nonumber\\
sb^a&=&0\;,\label{4a}
\label{Q1}
\end{eqnarray}
where $s$ is the nilpotent BRST operator and $D^{ab}_{\mu}\;\equiv\;\delta^{ab}\partial_\mu -gf^{abc}A^c_\mu$ is the covariant derivative in the adjoint representation. Thus, the Yang-Mills action, together with the Dirac action, with the gauge fixed, has the form
\begin{eqnarray}
\Sigma_0&=&\Sigma_{YM}+\Sigma_{D}+\Sigma_{gf}\;,
\label{Q2}
\end{eqnarray}
where 
\begin{eqnarray}
\Sigma_{gf}&=&s\int d^4x\;\bar{c}^a\partial^{\mu}A^a_{\mu}=\int d^4x\left(b^a\partial^{\mu}A^a_{\mu}+\bar{c}^a\partial^\mu D^{ab}_\mu c^b\right)\;
\label{Q3}
\end{eqnarray}
is the gauge fixing action. In Table \ref{table1} the quantum numbers of the fields and background vectors are presented.

\begin{table}[h]
\centering
\begin{tabular}{|c|c|c|c|c|c|c|c|c|}
	\hline 
quantities & $A$ & $b$ &$c$ & $\overline{c}$ & $\psi$ & $\overline{\psi}$ & $v$ & $b$ \\
	\hline 
UV dimension & $1$ & $2$ & $0$ & $2$ & $3/2$& $3/2$ & 1 & 1\\ 
Ghost number & $0$ & $0$ & $1$& $-1$& $0$& $0$ & 0 & 0 \\ 
Spinor number & $0$ & $0$ & $0$& $0$& $1$&$-1 $ & 0 & 0 \\ 
Statistics & $0$& $0$ & $1$ & $-1$& $1$ & $-1$ & 0 & 0 \\ 
\hline 
\end{tabular}
\caption{Quantum numbers of the fields and background tensors.}
\label{table1}
\end{table}

\subsection{BRST embedding of the sources}

As pointed by Symanzik in \cite{Symanzik:1969ek}, we must be careful with the renormalization of theories presenting explicit symmetry breaking. In fact, if a symmetry of a classical theory is explicitly violated, nothing will avoid, without a breaking control, that this explicit breaking becomes worse at high orders in perturbation theory, plaguing the model with non-physical vertices and modes. In the present model Lorentz symmetry is explicitly broken, with breaking characterized by the constant background fields $v^\mu$ and $\kappa^\mu$, in the bosonic and fermionic sector, respectively. Moreover, in the bosonic sector, we have a soft gauge symmetry breaking, i.e., proportional to the background field $v^\mu$. Nevertheless, whether we consider that such a vector has null curl, or simply that its derivative is vanishing, the gauge symmetry is preserved. Note, however, that this consists into assume one-shell gauge symmetry. However, for the sake of generality and in order to be more rigorous, the algebraic formalism consists in to use off-shell symmetry of the Ward identities. In fact, even in the usual Yang-Mills theory, after the gauge fixing, the gauge symmetry is preserved only on-shell. The BRST formalism allows a symmetry that close off-shell: the BRST symmetry. Moreover, the study of the quantum theory stay restricted to a cohomology problem (see Sects.~\ref{GA} and \ref{QS}). Note that, essentially, the Symanzik method consists into embedding the theory that contains broken symmetries into a larger theory, without breaks, through introduction of external sources.

Following this procedure, in combination with the BRST quantization, we restore the Lorentz symmetry and gain off-shell BRTS symmetry. In fact, as performed before in \cite{Santos:2016bqc,Santos:2015koa,Santos:2014lfa}, the background $\kappa^\mu$, which is coupled to a BRST invariant composite operator, it is replaced by a local external BRST invariant source $B^\mu$
\begin{equation}
sB^{\mu}=0\;.
\label{Q4}
\end{equation}
In the bosonic sector, once the background field $v^\mu$ is coupled to a BRST non-invariant composite operator, we substitute it by two local external sources $J^{\mu\nu\alpha}$ and $\eta^{\mu\nu\alpha}$ and their respective BRST complements $\lambda^{\mu\nu\alpha}$ and $\tau^{\mu\nu\alpha}$, namely
\begin{eqnarray}
s\lambda_{\mu\nu\alpha}&=&J_{\mu\nu\alpha}\;,\nonumber\\
sJ_{\mu\nu\alpha}&=&0\;,\nonumber\\
s\eta_{\mu\nu\alpha}&=&\tau_{\mu\nu\alpha}\;,\nonumber\\
s\tau_{\mu\nu\alpha}&=&0\;.
\label{Q5}
\end{eqnarray}
Once the renormalizability is studied (It is worthwhile to understand that the complete analysis of the renormalizability is performed within the action that presents Lorentz, CPT and BRST symmetries \cite{Santos:2016bqc,Santos:2015koa,Santos:2014lfa} and not the physical one.), the physical action \eqref{YM0} is recovered when the sources attain their physical values\footnote{See many examples of the method at \cite{Santos:2016bqc,Santos:2015koa,Symanzik:1969ek,Santos:2014lfa,Zwanziger:1992qr,Dudal:2005na,Baulieu:2008fy,Baulieu:2009xr,Dudal:2011gd,Pereira:2013aza}.},
\begin{eqnarray}
J_{\mu\nu\alpha}\mid_{phys}&=&\tau_{\mu\nu\alpha}\mid_{phys}\;=\;v^{\beta}\epsilon_{\beta\mu\nu\alpha}\;,\nonumber\\
\lambda_{\mu\nu\alpha}\mid_{phys}&=&\eta_{\mu\nu\alpha}\mid_{phys}\;=\;0\;,\nonumber\\
B^{\mu}\mid_{phys}&=&\kappa^{\mu}\;.
\label{Q6}
\end{eqnarray}
Hence, the Symanzik procedure implies that the bosonic sector will have the form
\begin{eqnarray}
\Sigma_{B}&=&s\int d^4x\left(\lambda^{\mu\nu\alpha}A^a_{\mu}\partial_{\nu}A^a_{\alpha}+\frac{1}{3}\eta^{\mu\nu\alpha}gf^{abc}A^a_{\mu}A^b_{\nu}A^c_{\alpha}\right)\nonumber\\
&=&\int d^4x\left[J^{\mu\nu\alpha}A^a_{\mu}\partial_{\nu}A^a_{\alpha}+\frac{1}{3}\tau^{\mu\nu\alpha}gf^{abc}A^a_{\mu}A^b_{\nu}A^c_{\alpha}+\lambda^{\mu\nu\alpha}\partial_{\mu}c^a\partial_{\nu}A^a_{\alpha}+\right.\nonumber\\
&+&\left.(\eta^{\mu\nu\alpha}-\lambda^{\mu\nu\alpha})gf^{abc}A^a_{\mu}A^b_{\nu}\partial_{\alpha}c^c\right]\;,
\label{Q7}
\end{eqnarray}
and the fermionic sector will have the form
\begin{eqnarray}
\Sigma_{F}&=&-\int d^4x\;B^{\mu}\overline{\psi}\gamma_5\gamma_{\mu}\psi\;.
\label{Q8}
\end{eqnarray}

From power-counting inspection, it is possible to consider one more term, which is a coupling between quadratic local composite operators and the external sources, namely,
\begin{align}
\Sigma_{LCO}&=s\int d^4x\left[\left(\alpha_1\lambda^{\mu\nu\alpha}J_{\mu\nu\alpha}+\alpha_2\lambda^{\mu\nu\alpha}\tau_{\mu\nu\alpha}+\alpha_3\eta^{\mu\nu\alpha}J_{\mu\nu\alpha}+\alpha_4\eta^{\mu\nu\alpha}\tau_{\mu\nu\alpha}\right)\frac{1}{2}A^a_\beta A^{a\beta}
+\right.\displaybreak[3]\nonumber\\
&+\left.\left(\beta_1\lambda^{\mu\alpha\beta}J_{\nu\alpha\beta}+\beta_2\lambda^{\mu\alpha\beta}\tau_{\nu\alpha\beta}+\beta_3\eta^{\mu\alpha\beta}J_{\nu\alpha\beta}+\beta_4\eta^{\mu\alpha\beta}\tau_{\nu\alpha\beta}\right)A^a_\mu A^{a\nu}
\right]\;\displaybreak[3]\nonumber\\
&=\int d^4x\left[\left(\alpha_1J^{\mu\nu\alpha}J_{\mu\nu\alpha}+\alpha_2J^{\mu\nu\alpha}\tau_{\mu\nu\alpha}+\alpha_3\tau^{\mu\nu\alpha}J_{\mu\nu\alpha}+\alpha_4\tau^{\mu\nu\alpha}\tau_{\mu\nu\alpha}\right)\frac{1}{2}A^a_\beta A^{a\beta}
+\right.\displaybreak[3]\nonumber\\
&+\left.\left(\beta_1J^{\mu\alpha\beta}J_{\nu\alpha\beta}+\beta_2J^{\mu\alpha\beta}\tau_{\nu\alpha\beta}+\beta_3\tau^{\mu\alpha\beta}J_{\nu\alpha\beta}+\beta_4\tau^{\mu\alpha\beta}\tau_{\nu\alpha\beta}\right)A^a_\mu A^{a\nu}
+\right.\displaybreak[3]\nonumber\\
&+\left.
\left(\alpha_1\lambda^{\mu\nu\alpha}J_{\mu\nu\alpha}+\alpha_2\lambda^{\mu\nu\alpha}\tau_{\mu\nu\alpha}+\alpha_3\eta^{\mu\nu\alpha}J_{\mu\nu\alpha}+\alpha_4\eta^{\mu\nu\alpha}\tau_{\mu\nu\alpha}\right)A^a_\beta \partial^\beta c^a
+\right.\displaybreak[3]\nonumber\\
&+\left.\left(\beta_1\lambda^{\mu\alpha\beta}J_{\nu\alpha\beta}+\beta_2\lambda^{\mu\alpha\beta}\tau_{\nu\alpha\beta}+\beta_3\eta^{\mu\alpha\beta}J_{\nu\alpha\beta}+\beta_4\eta^{\mu\alpha\beta}\tau_{\nu\alpha\beta}\right)(A^a_\mu \partial^\nu c^a+\partial_\mu c^aA^a_\nu)\right]\;.
\label{Q10}
\end{align}

Finally, we introduce a set of BRST invariant sources $\Omega^a_{\mu},\;L^a,\;\overline{Y}$ and $Y$, in order to control the nonlinear BRST transformations of the quantum fields. Thus, we have one more term to consider,
\begin{eqnarray}
\Sigma_{ext}&=&\int d^4x\left(\Omega^{a\mu}sA^a_{\mu}+L^asc^a+\overline{Y}s\psi-s\overline{\psi}Y\right)\nonumber\\
&=&\int d^4x\left(-\Omega^{a\mu}D^{ab}_{\mu}c^b+\frac{g}{2}f^{abc}L^ac^bc^c+g\overline{Y}c^aT^a\psi-g\overline{\psi}c^aT^aY\right)\;.
\label{Q9}
\end{eqnarray}
In Table \ref{table3} we present the quantum numbers of all sources we have defined. 
\begin{table}[h]
\centering
\begin{tabular}{|c|c|c|c|c|c|c|c|c|c|}
	\hline 
sources & $Y$ & $\overline{Y}$ & $\Omega$ & $L$ &$\lambda$ & $J$ & $\eta$ & $\tau$& $B$ \\
	\hline 
UV dimension & $5/2$ & $5/2$ & $2$ &  $4$& $1$ & $1$ &  $1$ & $1$ & $1$ \\ 
Ghost number & $-1$ & $-1$ & $-1$ &  $-2$ & $-1$ & $0$ &$-1$ & $0$ & $0$ \\ 
Spinor number & $1$ & $-1$ & $0$ &  $0$ & $0$ & $0$ & $0$ & $0$ & $0$ \\ 
Statistics &  $0$ & $-2$ &  $-1$ &  $-2$ & $-1$ & $0$ & $-1$ & $0$ & $0$  \\ 
\hline 
\end{tabular}
\caption{Quantum numbers of the sources.}
\label{table3}
\end{table}

It is not difficult to see, still from power-counting analysis, that an action depending only on the external sources can also be included, i.e., a vacuum term. However, this term does not affect the dynamical content of the theory and, for simplicity, we omit it, see \cite{Santos:2014lfa}. Thus, the most general action to be considered is given by
\begin{eqnarray}
\Sigma&=&\Sigma_{YM}+\Sigma_D+\Sigma_{gf}+\Sigma_{B}+\Sigma_{F}+\Sigma_{LCO}\;.
\label{Q11}
\end{eqnarray}
It is straightforward to show that this action is BRST invariant.

With this procedure, the Lorentz and BRST symmetries are well established and the renormalizability study can be safely done \cite{Santos:2014lfa}.

\subsection{Ward identities}\label{WI}

The symmetries enjoyed by the action \eqref{Q11} are functionally represented by the following set of Ward identities:

\begin{itemize}
	\item Slavnov-Taylor identity
	\begin{eqnarray}
\mathcal{S}(\Sigma)&\equiv&\int d^4x\left(\frac{\delta \Sigma}{\delta \Omega^{a\mu}}\frac{\delta \Sigma}{\delta A^a_{\mu}}+\frac{\delta \Sigma}{\delta L^a}\frac{\delta \Sigma}{\delta c^a}+\frac{\delta \Sigma}{\delta \overline{Y}}\frac{\delta \Sigma}{\delta \psi}-\frac{\delta \Sigma}{\delta Y}\frac{\delta \Sigma}{\delta \overline{\psi}}+b^a\frac{\delta \Sigma}{\delta \bar{c}^a}+\right.\nonumber\\
&+&\left.J_{\mu\nu\alpha}\frac{\delta \Sigma}{\delta\lambda_{\mu\nu\alpha}}+\tau_{\mu\nu\alpha}\frac{\delta \Sigma}{\delta\eta_{\mu\nu\alpha}}\right)=0\;.
\label{WI1}
\end{eqnarray}

\item Gauge fixing and antighost equations
\begin{eqnarray}
\frac{\delta \Sigma}{\delta b^a}&=&\partial^{\mu}A^a_{\mu}\;,\nonumber\\
\frac{\delta \Sigma}{\delta \bar{c}^a}+\partial_\mu \frac{\delta \Sigma}{\delta \Omega^a_\mu }&=&0\;.
\label{WI2}
\end{eqnarray}

\item Ghost equation
\begin{eqnarray}
\mathcal{G}^a\Sigma&=&\Delta^a_{cl}\;,
\label{WI3}
\end{eqnarray}
with 
\begin{eqnarray}
\mathcal{G}^a&=&\int d^4x\left(\frac{\delta}{\delta c^a}+gf^{abc}\bar{c}^b\frac{\delta}{\delta b^c}\right)\;,
\label{WI3a}
\end{eqnarray}
and
\begin{eqnarray}
\Delta^a_{cl}&=&\int d^4x\left[gf^{abc}\left(\Omega^b_{\mu}A^c_{\mu}-L^bc^c\right)+g\overline{Y}T^a\psi+g\overline{\psi}T^aY\right]\;.
\label{WI3b}
\end{eqnarray}
\end{itemize}
Since the breaking at Ward identities \eqref{WI2} and \eqref{WI3} are linear in the fields, they will remain at classical level \cite{Piguet:1995er}. 

For future purposes, let us define $\mathcal{F}$, a general functional with even ghost number. The Slavnov-Taylor operator acting on $\mathcal{F}$ is denoted by
\begin{eqnarray}
\mathcal{S}(\mathcal{F})&\equiv&\int d^4x\left(\frac{\delta\mathcal{F}}{\delta \Omega^{a\mu}}\frac{\delta\mathcal{F}}{\delta A^a_{\mu}}+\frac{\delta \mathcal{F}}{\delta L^a}\frac{\delta\mathcal{F}}{\delta c^a}+\frac{\delta \mathcal{F}}{\delta \overline{Y}}\frac{\delta\mathcal{F}}{\delta \psi}-\frac{\delta\mathcal{F}}{\delta Y}\frac{\delta\mathcal{F}}{\delta \overline{\psi}}+b^a\frac{\delta\mathcal{F}}{\delta \bar{c}^a}+\right.\nonumber\\
&+&\left.J_{\mu\nu\alpha}\frac{\delta\mathcal{F}}{\delta\lambda_{\mu\nu\alpha}}+\tau_{\mu\nu\alpha}\frac{\delta \mathcal{F}}{\delta\eta_{\mu\nu\alpha}}\right)\;.
\label{WI4}
\end{eqnarray}
We can define the linearized Slavnov-Taylor operator as
\begin{eqnarray}
\mathcal{S}_{\mathcal{F}}&\equiv&\int d^4x\left(\frac{\delta\mathcal{F}}{\delta \Omega^{a\mu}}\frac{\delta}{\delta A^a_{\mu}}+\frac{\delta\mathcal{F}}{\delta A^a_{\mu}}\frac{\delta}{\delta \Omega^{a\mu}}+\frac{\delta \mathcal{F}}{\delta L^a}\frac{\delta}{\delta c^a}+\frac{\delta\mathcal{F}}{\delta c^a}\frac{\delta }{\delta L^a}+\frac{\delta \mathcal{F}}{\delta \overline{Y}}\frac{\delta }{\delta \psi}+\frac{\delta \mathcal{F}}{\delta\psi}\frac{\delta}{\delta\overline{Y}}+\right.\nonumber\\
&-&\left.\frac{\delta \mathcal{F}}{\delta Y}\frac{\delta}{\delta \overline{\psi}}-\frac{\delta \mathcal{F}}{\delta\overline{\psi}}\frac{\delta}{\delta Y}+b^a\frac{\delta}{\delta \overline{c}^a}+J_{\mu\nu\alpha}\frac{\delta }{\delta\lambda_{\mu\nu\alpha}}+\tau_{\mu\nu\alpha}\frac{\delta }{\delta\eta_{\mu\nu\alpha}}\right)\;.
\label{WI5}
\end{eqnarray}
The following identities hold
\begin{eqnarray}
\mathcal{S}_{\mathcal{F}}\mathcal{S}(\mathcal{F})&=&0\;,\;\;\forall\;\;\mathcal{F}\;,\nonumber\\
\mathcal{S}_{\mathcal{F}}\mathcal{S}_{\mathcal{F}}&=&0\;,\;\;\textrm{if}\;\;\mathcal{S}(\mathcal{F})\;=\;0\;.
\label{WI6} 
\end{eqnarray}

\section{Gauge anomalies}\label{GA}

In order to analyze the renormalizability of the model described by the action \eqref{Q11}, we need to prove that (\textit{i}) the Ward identities \eqref{WI1}--\eqref{WI3} are not anomalous at quantum level, and (\textit{ii}) that the action is stable at quantum level. It is known that there is no room for gauge anomalies in the pure Yang-Mills theory with Lorentz violation \cite{Piguet:1995er,Santos:2014lfa}. Here, however, there are fermions and an additional Lorentz-breaking sector. Following \cite{Santos:2016bqc,Piguet:1995er}, is a trivial exercise to check that the Ward identities \eqref{WI2} and \eqref{WI3} are not anomalous at quantum level, i.e., 
\begin{itemize}
	
\item Gauge fixing and anti-ghost equations
\begin{eqnarray}
\frac{\delta \Gamma}{\delta b^a}&=&\partial^{\mu}A^a_{\mu}\;,\nonumber\\
\frac{\delta \Gamma}{\delta \bar{c}^a}+\partial_\mu \frac{\delta \Gamma}{\delta \Omega^a_\mu }&=&0\;.
\label{GA1}
\end{eqnarray}

\item Ghost equation
\begin{eqnarray}
\mathcal{G}^a\Gamma&=&\Delta^a_{cl}\;,
\label{GA2}
\end{eqnarray}
\end{itemize}
where $\Gamma$ stands for the quantum action, namely,
\begin{eqnarray}
\Gamma&=&\sum_{n=0}^{\infty}\hbar^n\Gamma^{(n)}\;\;\textrm{with}\;\;\Gamma^{(0)}\;=\;\Sigma\;.
\label{GA2a}
\end{eqnarray}
Here we shall show that the Ward identity \eqref{WI1} also is true for the action $\Gamma$. In fact, this is the main Ward identity of the model and if this identity is ruined, renormalizability is lost. In order to do that, using the quantum action principle (QAP), we assume that such Ward identity breaks down at order $\hbar^n$ in perturbation theory, as follows
\begin{eqnarray}
\mathcal{S}(\Gamma)&=&\hbar^n\Delta^{(1)}+\mathcal{O}(\hbar^{n+1})\;,
\label{GA3}
\end{eqnarray}
were $\Delta^{(1)}$ is a local integrated polynomial in the fields and external sources, of ghost number one and dimension bounded by four. From identity \eqref{WI6}, we get
\begin{eqnarray}
\mathcal{S}_{\Gamma}\Delta^{(1)}&=&0\;.
\label{GA4}
\end{eqnarray}
This identity is the so-called Wess-Zumino consistence condition for the anomaly \cite{Wess:1971yu}. The Eq.~\eqref{GA4} defines a cohomology problem in the space of the integrated local polynomial on the fields and external sources of ghost number one and dimension bounded by four. The most general solution for \eqref{GA4} has the form
\begin{eqnarray}
\Delta^{(1)}&=&r\mathcal{A}+\mathcal{S}_{\Gamma}\hat{\Delta}^{(0)}\;,
\label{GA5}
\end{eqnarray}
were $\mathcal{A}$ is a local polynomial in the fields and sources and $r$ is an arbitrary parameter. This parameter is not determined by algebraic methods: only an explicit computation of Feynman diagrams can determine it. Anomalies only are present when there exist nontrivial solutions, i.e., $\mathcal{A}\neq\mathcal{S}_{\Gamma}\hat{\mathcal{A}}$. In fact, in this case the Slavnov-Taylor operator only can be implemented to $\hbar^{n-1}$ order in perturbation theory, and just the trivial part can be reabsorbed by the introduction of the noninvariant counterterm $-\hat{\mathcal{A}}$ into the classical action. A direct consequence of this is that trivial solutions for the cohomology problem always can be eliminated, implying in the anomaly-freedom of the model. Furthermore, if the $r$ parameter can be made to vanish, through sum on all species of fermions in a family, for instance, the anomaly also can be eliminated. A nonrenormalization theorem can assure this property to all orders in perturbation theory \cite{Adler:1969er}.

Inhere, the most general solution $\Delta^{(1)}$ must take into account the following criteria: dimension bounded by four, ghost number one, polynomial on the fields and sources, Lorentz, C, P and T invariant -- considering that in this stage these symmetries are restored. However, it is known that such a solution will not depend on the fermion fields and on the external sources (since they form BRST doublets), i.e., their contribution for the anomaly are trivial \cite{Santos:2016bqc,Piguet:1995er}. In fact, this last result is general \cite{Barnich:1994ve,Barnich:1994db,Barnich:1994mt}. In fact, supposing for now that the breaking happens at $\hbar$ order in perturbation theory, the other terms that could appear from usual Dirac-Yang-Mills theory also do not contribute, see \cite{Piguet:1995er}. Thus, the remaining term could depend only on $A$, $c$ and $B$, namely $\Delta^{(1)}\equiv\Delta^{(1)}(A,c,B)$. Hence, the linearized Slavnov-Taylor operator $\mathcal{S}_{\Gamma}$ can be identified with the BRST operator, $s$, because the action of $\mathcal{S}_{\Gamma}$ on $(A,c,B)$ is the same as $s$. Then, the problem \eqref{GA5} is reduced to the simpler cohomology problem given by 
\begin{eqnarray}
s\Delta^{(1)}(A,c,B)&=&0\;.
\label{GA6}
\end{eqnarray}
It is possible to see that there is only one term that satisfy the Eq.~\eqref{GA6}. Thus, the most general solution for the anomaly reads
\begin{eqnarray}
\Delta^{(1)}&=&r\int d^4x\;\epsilon_{\mu\nu\alpha\beta}B^{\mu}\partial^{\nu}c^a\partial^{\alpha}A^{a\beta}\;.
\label{GA7}
\end{eqnarray}
However, it is straightforward to show that this term can be written as $\Delta^{(1)}=s\Delta^{(0)}$, where
\begin{eqnarray}
\Delta^{(0)}&=&-r\int d^4x\;\epsilon_{\mu\nu\alpha\beta}B^{\mu}\left(A^{a\nu}\partial^{\alpha}A^{a\beta}+\frac{g}{3}f^{abc}A^{a\nu}A^{b\alpha}A^{c\beta}\right)\;.
\label{GA8}
\end{eqnarray}
This means that there is no non-trivial solution for the Eq.~\eqref{GA6} (or Eq.~\eqref{GA4}). Thus such anomaly can be eliminated by introduction of the noninvariant counterterm $-\Delta^{(0)}$ into the classical action. Aftermath, the model described by the action \eqref{Q11} is anomaly-free at first order in perturbation theory. Moreover, since the method is recursive, this property remains at all orders in perturbation theory.

\section{Stability}\label{QS}

Once we have shown that the Ward identities \eqref{WI1}--\eqref{WI3} are not anomalous at quantum level, we can now study the quantum stability of the model \eqref{Q11}, i.e., to seek for the most general invariant counterterm, $\Sigma^{ct}$, which can be freely added to the classical action $\Sigma$ at any order in perturbation theory. Such a counterterm must have dimension bounded by four and vanishing ghost number and also must obey the following constraints
\begin{subequations}
\begin{eqnarray}
\label{QS1a}
\mathcal{S}_{\Sigma}\Sigma^{ct}&=&0\;,\\
\label{QS1b}
\frac{\delta \Sigma^{ct}}{\delta b^a}&=&0\;,\\
\label{QS1c}
\left(\frac{\delta }{\delta \bar{c}^a}+\partial_\mu \frac{\delta }{\delta \Omega^a_\mu }\right)\Sigma^{ct}&=&0\;,\\
\label{QS1d}
\mathcal{G}^a\Sigma^{ct}&=&0\;,
\end{eqnarray}
\end{subequations}
where $\mathcal{S}_{\Sigma}$ linearized nilpotent Slavnov-Taylor operator is given by \eqref{QS1a}
\begin{eqnarray}
\mathcal{S}_{\Sigma}&=&\int d^4x\left(\frac{\delta \Sigma}{\delta \Omega^a_{\mu}}\frac{\delta}{\delta A^{a\mu}}+\frac{\delta \Sigma}{\delta A^{a\mu}}\frac{\delta }{\delta \Omega^a_{\mu}}+\frac{\delta \Sigma}{\delta L^a}\frac{\delta }{\delta c^a}+\frac{\delta \Sigma}{\delta  c^a}\frac{\delta}{\delta L^a}+\frac{\delta \Sigma}{\delta \overline{Y}}\frac{\delta }{\delta \psi}+\frac{\delta \Sigma}{\delta\psi}\frac{\delta}{\delta\overline{Y}}-\frac{\delta \Sigma}{\delta Y}\frac{\delta}{\delta \overline{\psi}}\right.\nonumber\\
&-&\left.\frac{\delta \Sigma}{\delta\overline{\psi}}\frac{\delta}{\delta Y}+b^a\frac{\delta}{\delta \bar{c}^a}+J_{\mu\nu\alpha}\frac{\delta }{\delta\lambda_{\mu\nu\alpha}}+\tau_{\mu\nu\alpha}\frac{\delta }{\delta\eta_{\mu\nu\alpha}}\right)\;.
\label{QS2}
\end{eqnarray}
The constraint \eqref{QS1a} identifies the invariant counterterm as the solution of the cohomology problem for the operator $\mathcal{S}_{\Sigma}$ in the space of the integrated local field polynomials of dimension four and vanishing ghost number. From the general results of cohomology, it follows that $\Sigma^{ct}$ can be written as \cite{Piguet:1995er}
\begin{eqnarray}
\Sigma^{ct}&=&-\frac{1}{4}\int d^4x\;a_0F^a_{\mu\nu}F^{a\mu\nu}+\int d^4x\left(a_1i\overline{\psi}\gamma^{\mu}D_{\mu}\psi-a_2m\overline{\psi}\psi-a_3B^{\mu}\overline{\psi}\gamma_5\gamma_{\mu}\psi\right)+\nonumber\\
&+&\mathcal{S}_{\Sigma}\Delta^{(-1)}\;,
\label{QS3}
\end{eqnarray}
where $\Delta^{(-1)}$ is the most general local polynomial counterterm with dimension bounded by four and ghost number $-1$, given by
\begin{eqnarray}
\Delta^{(-1)}&=&\int d^4x\left[a_4\Omega^{a\mu}A^a_{\mu}+a_5\partial^{\mu}\bar{c}^aA^a_{\mu}+a_6L^ac^a+\frac{a_7}{2}\bar{c}^ab^a+a_8\frac{g}{2}f^{abc}\bar{c}^a\bar{c}^bc^c+a_9\overline{Y}\psi+\right.\nonumber\\
&+&\left.a_{10}\overline{\psi}Y+\left(a_{11}\lambda^{\mu\nu\alpha}+a_{12}\eta^{\mu\nu\alpha}\right)A^a_{\mu}\partial_{\nu}A^a_{\alpha}+\left(a_{13}\lambda^{\mu\nu\alpha}+a_{14}\eta^{\mu\nu\alpha}\right)\frac{g}{3}f^{abc}A^a_{\mu}A^b_{\nu}A^c_{\alpha}+\right.\nonumber\\
&+&\left.\left(a_{15}\lambda_{\alpha\beta\gamma}+a_{16}\eta_{\alpha\beta\gamma}\right)\epsilon^{\alpha\beta\gamma\mu}\overline{\psi}\gamma_5\gamma_{\mu}\psi+\right.\nonumber\\
&+&\left.\left(a_{17}\alpha_1\lambda^{\mu\nu\alpha}J_{\mu\nu\alpha}+a_{18}\alpha_2\lambda^{\mu\nu\alpha}\tau_{\mu\nu\alpha}+a_{19}\alpha_3\eta^{\mu\nu\alpha}J_{\mu\nu\alpha}+a_{20}\alpha_4\eta^{\mu\nu\alpha}\tau_{\mu\nu\alpha}\right)\frac{1}{2}A^a_\beta A^{a\beta}
+\right.\nonumber\\
&+&\left.\left(a_{21}\beta_1\lambda^{\mu\alpha\beta}J_{\nu\alpha\beta}+a_{22}\beta_2\lambda^{\mu\alpha\beta}\tau_{\nu\alpha\beta}+a_{23}\beta_3\eta^{\mu\alpha\beta}J_{\nu\alpha\beta}+a_{24}\beta_4\eta^{\mu\alpha\beta}\tau_{\nu\alpha\beta}\right)A^a_\mu A^{a\nu}
\right]\;.\nonumber\\
\label{QS4}
\end{eqnarray}
From Eq.~\eqref{QS1b} one finds that $a_4=a_5$ and $a_8=a_7=0$. Moreover, from Eq.~\eqref{QS1d} one finds that $a_6=0$. Thus, the form of the most general counterterm allowed by the Ward identities is given by
\begin{align}
\Sigma^{ct}&=-\frac{1}{4}\int d^4x\;a_0F^a_{\mu\nu}F^{a\mu\nu}+\int d^4x\left[\left(a_{10}-a_9+a_1\right)i\overline{\psi}\gamma^{\mu}D_{\mu}\psi-\left(a_{10}-a_9+a_2\right)m\overline{\psi}\psi\right]+\displaybreak[3]\nonumber\\
&+a_4\int d^4x\left[A^a_\mu\frac{\delta \Sigma_{YM}}{\delta A^a_\mu }+A^a_\mu\frac{\delta \Sigma_{B}}{\delta A^a_\mu }+(\Omega^a_{\mu}+\partial_{\mu}\bar{c}^a)\partial^{\mu}c^a+ig\overline{\psi}\gamma^{\mu}\psi A^a_{\mu}T^a+\right.\displaybreak[3]\nonumber\\
&+\left.\left(\alpha_1J^{\mu\nu\alpha}J_{\mu\nu\alpha}+\alpha_2J^{\mu\nu\alpha}\tau_{\mu\nu\alpha}+\alpha_3\tau^{\mu\nu\alpha}J_{\mu\nu\alpha}+\alpha_4\tau^{\mu\nu\alpha}\tau_{\mu\nu\alpha}\right)A^a_\beta A^{a\beta}
+\right.\displaybreak[3]\nonumber\\
&+\left.2\left(\beta_1J^{\mu\alpha\beta}J_{\nu\alpha\beta}+\beta_2J^{\mu\alpha\beta}\tau_{\nu\alpha\beta}+\beta_3\tau^{\mu\alpha\beta}J_{\nu\alpha\beta}+\beta_4\tau^{\mu\alpha\beta}\tau_{\nu\alpha\beta}\right)A^a_\mu A^{a\nu}
+\right.\displaybreak[3]\nonumber\\
&+\left.\left(\alpha_1\lambda^{\mu\nu\alpha}J_{\mu\nu\alpha}+\alpha_2\lambda^{\mu\nu\alpha}\tau_{\mu\nu\alpha}+\alpha_3\eta^{\mu\nu\alpha}J_{\mu\nu\alpha}+\alpha_4\eta^{\mu\nu\alpha}\tau_{\mu\nu\alpha}\right)A^a_\beta \partial^{\beta} c^a+\right.\displaybreak[3]\nonumber\\
&+\left.\left(\beta_1\lambda^{\mu\alpha\beta}J_{\nu\alpha\beta}+\beta_2\lambda^{\mu\alpha\beta}\tau_{\nu\alpha\beta}+\beta_3\eta^{\mu\alpha\beta}J_{\nu\alpha\beta}+\beta_4\eta^{\mu\alpha\beta}\tau_{\nu\alpha\beta}\right)(A^a_\mu \partial^\nu c^a+\partial_\mu c^aA^{a\nu})\right]+\displaybreak[3]\nonumber\\
&+\int d^4x\left[J^{\mu\nu\alpha}\left(a_{11}A^a_{\mu}\partial_{\nu}A^a_{\alpha}+a_{13}\frac{g}{3}f^{abc}A^a_{\mu}A^b_{\nu}A^c_{\alpha}\right)+a_{11}\lambda^{\mu\nu\alpha}\partial_{\mu}c^a\partial_{\nu}A^a_{\alpha}+\right.\displaybreak[3]\nonumber\\
&+\left.\tau^{\mu\nu\alpha}\left(a_{12}A^a_{\mu}\partial_{\nu}A^a_{\alpha}+a_{14}\frac{g}{3}f^{abc}A^a_{\mu}A^b_{\nu}A^c_{\alpha}\right)+a_{12}\eta^{\mu\nu\alpha}\partial_{\mu}c^a\partial_{\nu}A^a_{\alpha}+\right.\displaybreak[3]\nonumber\\
&+\left.
(a_{13}-a_{11})\lambda^{\mu\nu\alpha}gf^{abc}A^a_\mu A^c_\alpha \partial_\nu c^b+(a_{14}-a_{12})\eta^{\mu\nu\alpha}gf^{abc}A^a_\mu A^c_\alpha \partial_\nu c^b+
\right.\displaybreak[3]\nonumber\\
&+\left.\left(a_{17}\alpha_1J^{\mu\nu\alpha}J_{\mu\nu\alpha}+a_{18}\alpha_2J^{\mu\nu\alpha}\tau_{\mu\nu\alpha}+a_{19}\alpha_3\tau^{\mu\nu\alpha}J_{\mu\nu\alpha}+a_{20}\alpha_4\tau^{\mu\nu\alpha}\tau_{\mu\nu\alpha}\right)\frac{1}{2}A^a_\beta A^{a\beta}
+\right.\displaybreak[3]\nonumber\\
&+\left.\left(a_{17}\alpha_1\lambda^{\mu\nu\alpha}J_{\mu\nu\alpha}+a_{18}\alpha_2\lambda^{\mu\nu\alpha}\tau_{\mu\nu\alpha}+a_{19}\alpha_3\eta^{\mu\nu\alpha}J_{\mu\nu\alpha}+a_{20}\alpha_4\eta^{\mu\nu\alpha}\tau_{\mu\nu\alpha}\right)A^a_\beta \partial^\beta
c^a+\right.\displaybreak[3]\nonumber\\
&+\left.\left(a_{21}\beta_1J^{\mu\alpha\beta}J_{\nu\alpha\beta}+a_{22}\beta_2J^{\mu\alpha\beta}\tau_{\nu\alpha\beta}+a_{23}\beta_3\tau^{\mu\alpha\beta}J_{\nu\alpha\beta}+a_{24}\beta_4\tau^{\mu\alpha\beta}\tau_{\nu\alpha\beta}\right)A^a_\mu A^{a\nu}
+\right.\displaybreak[3]\nonumber\\
&+\left.\left(a_{21}\beta_1\lambda^{\mu\alpha\beta}J_{\nu\alpha\beta}+a_{22}\beta_2\lambda^{\mu\alpha\beta}\tau_{\nu\alpha\beta}+a_{23}\beta_3\eta^{\mu\alpha\beta}J_{\nu\alpha\beta}+a_{24}\beta_4\eta^{\mu\alpha\beta}\tau_{\nu\alpha\beta}\right)(A^a_\mu \partial^\nu c^a+\partial_\mu c^aA^{a\nu})\right]\displaybreak[3]\nonumber\\
&-\int d^4x\left[(a_{10}-a_9+a_3)B^{\mu}-a_{15}J_{\alpha\beta\gamma}\epsilon^{\alpha\beta\gamma\mu}-a_{16}\tau_{\alpha\beta\gamma}\epsilon^{\alpha\beta\gamma\mu}\right]\overline{\psi}\gamma_5\gamma_{\mu}\psi\;.
\label{31}
\end{align}

The last step in the  stability analysis is to infer if the counterterm $\Sigma^{ct}$ can be reabsorbed by the original action $\Sigma$ by means of the multiplicative redefinition of the fields, sources and parameters of the theory, according to
\begin{eqnarray}
\Sigma(\Phi,J,\xi)+\varepsilon\Sigma^{ct}(\Phi,J,\xi)&=&\Sigma(\Phi_0,J_0,\xi_0)+\mathcal{O}(\varepsilon^2)\;,
\label{32}
\end{eqnarray}
where $\varepsilon$ is a small perturbation parameter ($\hbar$ or the coupling parameter $g$) and the bare quantities are defined as
\begin{eqnarray}
\Phi_0&=&Z^{1/2}_{\Phi}\Phi\;,\;\;\;\;\;\;\Phi \in \left\{A,\;\overline{\psi},\;\psi,\;b,\;\overline{c},\;c\right\}\;,\nonumber\\
\mathcal{J}_0&=&Z_{\mathcal{J}}\mathcal{J}\;,\;\;\;\;\;\;\;\mathcal{J} \in \left\{\Omega,\;L,\overline{Y},\;Y,\;J,\;\lambda,\tau,\;\eta,\;B\right\}\;, \nonumber\\
\xi_0&=&Z_{\xi}\xi\;,\;\;\;\;\;\;\;\;\;\;\;\xi \in \left\{g,\;m\right\}\;.
\label{33}
\end{eqnarray}
Following this prescription, it is possible to check the renormalizability of the model, where the renormalization factors are given as follows: For the independent renormalization factors of the gauge field, coupling parameter, electron field and electron mass one finds
\begin{eqnarray}
Z_A^{1/2}&=&1+\varepsilon \left(\frac{a_0}{2}+a_4\right)\;,\nonumber\\
Z_g&=&1-\varepsilon\frac{a_0}{2}\;,\nonumber\\
Z_\psi^{1/2}&=&1+\varepsilon\frac{1}{2}\left(a_{10}-a_{9}+a_1\right)\;,\nonumber\\
Z_m&=&1+\varepsilon(a_2-a_1)\;,
\label{MR1}
\end{eqnarray}
while the renormalization factors of the ghosts, the Lautrup-Nakanishi field, $\Omega$,\;$L$ and $Y$ sources are not independent:
\begin{eqnarray}
Z_c&=&Z_{\bar{c}}\;\;=\;\;Z_A^{-1/2}Z_g^{-1}\;,\nonumber\\
Z_{\Omega}&=&Z_A^{-1/4}Z_g^{-1/2}\;,\nonumber\\
Z_L&=&Z_b^{-1/2}\;=\;Z_A^{1/2}\;,\nonumber\\
Z_Y&=&Z_{\overline{Y}}\;\;=\;\;Z_g^{-1/2}Z_A^{1/4}Z_{\psi}^{-1/2}\;.
\label{MR2}
\end{eqnarray}
At this point, we conclude that the renormalization properties of the usual Yang-Mills theory with fermions remain unchanged.

For the additional sector, once the sources $B^{\mu},\;J^{\alpha\beta\gamma}$ and $\;\tau^{\alpha\beta\gamma}$ share the same quantum numbers, matrix renormalization is required, i.e.,
\begin{equation}
\mathcal{J}_0=\mathcal{Z}_{\mathcal{J}}\mathcal{J}\;,\label{j}
\end{equation}
where $\mathcal{J}$ is a column matrix of sources that share the same quantum numbers. The quantity $Z_{\mathcal{J}}$ is a squared matrix with the associated renormalization factors. Thus
\begin{align}
\begin{pmatrix}
B_0^{\mu}\\
J_0^{\alpha\beta\gamma}\\
\tau_0^{\alpha\beta\gamma}
\end{pmatrix}&=\begin{pmatrix}
(Z_{BB})^{\mu}_{\phantom{\mu}\omega}& (Z_{BJ})^{\mu}_{\phantom{\mu}\lambda\rho\sigma}&(Z_{B\tau})^{\mu}_{\phantom{\mu}\lambda\rho\sigma}&\\
(Z_{JB})^{\alpha\beta\gamma}_{\phantom{\alpha\beta\gamma}\omega}&(Z_{JJ})^{\alpha\beta\gamma}_{\phantom{\alpha\beta\gamma}\lambda\rho\sigma}&(Z_{J\tau})^{\alpha\beta\gamma}_{\phantom{\alpha\beta\gamma}\lambda\rho\sigma}\\
(Z_{\tau B})^{\alpha\beta\gamma}_{\phantom{\alpha\beta\gamma}\omega}&(Z_{\tau J})^{\alpha\beta\gamma}_{\phantom{\alpha\beta\gamma}\lambda\rho\sigma}&(Z_{\tau\tau})^{\alpha\beta\gamma}_{\phantom{\alpha\beta\gamma}\lambda\rho\sigma}
\end{pmatrix}\begin{pmatrix}
B^{\omega}\\
J^{\lambda\rho\sigma}\\
\tau^{\lambda\rho\sigma}
\end{pmatrix}\;\nonumber\\
&=\begin{pmatrix}(1+\varepsilon(a_{3}-a_1))\delta^{\mu}_{\omega}&-\varepsilon a_{15} \epsilon_{\lambda\rho\sigma}^{\phantom{\lambda\rho\sigma}\mu}&-\varepsilon a_{16}\epsilon_{\lambda\rho\sigma}^{\phantom{\lambda\rho\sigma}\mu}\\
0&(1+\varepsilon (a_{11}-a_0))\delta^{\alpha}_{\lambda}\delta^{\beta}_{\rho}\delta^{\gamma}_{\sigma}&\varepsilon a_{12}\delta^{\alpha}_{\lambda}\delta^{\beta}_{\rho}\delta^{\gamma}_{\sigma}\\
0&\varepsilon a_{13} \delta^{\alpha}_{\lambda}\delta^{\beta}_{\rho}\delta^{\gamma}_{\sigma}&(1+\varepsilon (a_{14}-a_0))\delta^{\alpha}_{\lambda}\delta^{\beta}_{\rho}\delta^{\gamma}_{\sigma})
\end{pmatrix}\begin{pmatrix}
B^{\omega}\\
J^{\lambda\rho\sigma}\\
\tau^{\lambda\rho\sigma}
\end{pmatrix}\;.
\label{MR3}
\end{align}
The same rule will be used for the sources $\lambda_{\mu\nu\alpha}$ and $\eta_{\mu\nu\alpha}$, namely,
\begin{eqnarray}
\mathcal{J}=\begin{pmatrix}
\lambda_{\mu\nu\alpha}\\
\eta_{\mu\nu\alpha}
\end{pmatrix}\;&\mathrm{and}&\;\mathcal{Z}=\begin{pmatrix}
Z_{\lambda\lambda}& Z_{\lambda\eta}\\
Z_{\eta\lambda}& Z_{\eta\eta}
\end{pmatrix}\;,
\label{renm1x}
\end{eqnarray}
where we find
\begin{equation}
\mathcal{Z}=\mathbb{1}+\varepsilon\begin{pmatrix}
\frac{a_4}{2}-\frac{a_0}{2}+a_{11}& a_{12}\\
a_{13}&\frac{a_4}{2}-\frac{a_0}{2}+a_{14}
\end{pmatrix}\;.
\end{equation}
Finally, the renormalization factors of the dimensionless parameters read
\begin{align}
Z_{\alpha_1}&=1+\varepsilon\left(a_{17}-2a_{11}+a_0-\frac{\alpha_2+\alpha_3}{\alpha_1}a_{13}\right)\;,\displaybreak[3]\nonumber\\
Z_{\alpha_2}&=1+\varepsilon\left(a_{18}-a_{11}-a_{14}+a_0-\left(\frac{\alpha_1}{\alpha_2}a_{12}+\frac{\alpha_4}{\alpha_2}a_{13}\right)\right)\;,\displaybreak[3]\nonumber\\
Z_{\alpha_3}&=1+\varepsilon\left(a_{19}-a_{11}-a_{14}+a_0-\left(\frac{\alpha_1}{\alpha_3}a_{12}+\frac{\alpha_4}{\alpha_3}a_{13}\right)\right)\;,\displaybreak[3]\nonumber\\
Z_{\alpha_4}&=1+\varepsilon\left(a_{20}-2a_{14}+a_0-\frac{\alpha_2+\alpha_3}{\alpha_4}a_{12}\right)\;,\displaybreak[3]\nonumber\\
Z_{\beta_1}&=1+\varepsilon\left(a_{21}-2a_{11}+a_0-\frac{\beta_2+\beta_3}{\beta_1}a_{13}\right)\;,\displaybreak[3]\nonumber\\
Z_{\beta_2}&=1+\varepsilon\left(a_{22}-a_{11}-a_{14}+a_0-\left(\frac{\beta_1}{\beta_2}a_{12}+\frac{\beta_4}{\beta_2}a_{13}\right)\right)\;,\displaybreak[3]\nonumber\\
Z_{\beta_3}&=1+\varepsilon\left(a_{23}-a_{11}-a_{14}+a_0-\left(\frac{\beta_1}{\beta_3}a_{12}+\frac{\beta_4}{\beta_3}a_{13}\right)\right)\;,\displaybreak[3]\nonumber\\
Z_{\beta_4}&=1+\varepsilon\left(a_{24}-2a_{14}+a_0-\frac{\beta_2+\beta_3}{\beta_4}a_{12}\right)\;.
\label{ren7}
\end{align}

We concluded that the Lorentz-violating Yang-Mills theory with interacting fermions is stable at quantum level. More precisely, at all orders in perturbation theory.

\section{Conclusion}\label{FINAL}

In this work we have studied the issue of Chern-Simons-like term generation in the Lorentz-violating Yang-Mills theory with interacting fermions. For our proposes, we consider the usual non-Abelian Chern-Simons-like term and only one CPT-odd term in the fermionic sector: the one containing the background vector field $\kappa^\mu$. Since a Chern-Simons-like term could come from radiative corrections \cite{Gomes:2007rv}, the stability study of the model was needed. However, for such study, the Ward identities must remain true at quantum level, i.e., the model must be anomaly-free. Thus, the anomaly analysis was required to assure whether this property is true. In order to do that, we have employed the BRST quantization approach (once we are dealing with a gauge theory) in combination with the Symanzik method to control the breaking associated with the background fields. The algebraic renormalization technique gives us results which are independent of any renormalization scheme and are valid to all orders in perturbation theory. The results here found are: 
\begin{enumerate}
	\item The model here studied is anomaly-free, since there are no nontrivial solutions for the cohomology problem \eqref{GA4}. The Chern-Simons-like term appears at the trivial sector of the cohomology in the space of local polynomial of ghost number one. This means that: (\textit{i}) this term is redundant and (\textit{ii}) can be eliminated through renormalization conditions \cite{Piguet:1995er}.
	\item A Chern-Simons-like term is not generated by radiative corrections, as can be noted from the counterterm action \eqref{31}. This feature can be easily observed from renormalization factors shown at Eq.~\eqref{MR3}. The non generation of a Chern-Simons-like term is characterized by the fact that the source $J^{\mu\nu\alpha}$ does not receive quantum corrections from source $B^{\mu}$. In fact, since $J^{\mu\nu\alpha}$ belong to a BRST doublet, it cannot receive contributions from sources which are not at the trivial sector of the BRST cohomology. See \cite{Santos:2015koa} for more details. 
	\item Here we clarify why the (non)renormalization of the Chern-Simons-like action (see Eq.~\eqref{YM3}) or the Carroll-Field-Jackiw action (in the Abelian case) is not related to the (non)generation of a Chern-Simons-like action for non-Abelian or Abelian case. In fact, in the Abelian case, the Carroll-Field-Jackiw action does not renormalize and is not generated \cite{Santos:2015koa}. Inhere, we saw that the Chern-Simons-like action \eqref{YM3} does renormalize, as noted from the counterterm \eqref{31}. However, just like the Abelian case, a Chern-Simons-like term is not generated from radiative corrections since $J^{\mu\nu\alpha}$ does not receive contributions from $B^\mu$.
\end{enumerate}

It is worth to comment about a source of confusion. Although the model here studied presents background fields, the background field quantization method \cite{Dittrich:1985tr} is not employed. In the present model the background fields are inherent to the model. In the BRST quantization of the whole model the former background fields, which are not related to the quantum fields, were replaced by local external sources. In the case of the background field method, the background fields are counterparts for the usual quantum fields by means of that the latter are perturbations around the former. This point should be clear for the reader in order to distinguish the nice general results of \cite{Anselmi:2013kba,Barvinsky:2017zlx} from ours.

Another interesting point to be mentioned concern other classes of Lorentz-violating theories, for instance \cite{Anselmi:2007ri,Anselmi:2009ng,Anselmi:2010zh}, which also preserve renormalizability. In these theories the Lorentz symmetry breaking consists in to assume higher order space derivatives while the time derivatives remain at the same order as in the usual relativistic models. In these cases the renormalizability is assured by generalizing the usual power-counting analysis. From the weighted power-counting concept \cite{Anselmi:2013kba}, nonrenormalizable vertexes are put on a renormalizable form \cite{Anselmi:2009ng}. In the model here studied, however, the Lorentz violation manifests itself under particle Lorentz transformations, as the Carroll-Field-Jackiw models \cite{Carroll:1989vb}. Nevertheless, the usual Lorentz covariance -- space and time are treated on an equal footing -- is maintained.

\section*{Acknowledgements}

The Coordena\c c\~ao de Aperfei\c coamento de Pessoal de N\'ivel Superior (CAPES) is acknowledged for financial support.

\end{document}